\documentstyle    [ citesort]  {article}
\evensidemargin \oddsidemargin
\def \p {{\vec p}}
\def \k {{\vec k}}
\def \kap {{\vec \kappa}}
\def \H {{\cal H}}

\begin{document}
\baselineskip 20pt
\title{A Hamiltonian Formulation for Long Internal Waves}

\author{Yuri Lvov\footnote{Corresponding author, e-mail is lvovy$@$rpi.edu}\\
        Department of Mathematical Sciences\\
	Rensselaer Polytechnic Institute
\and
        Esteban G. Tabak\footnote{E-mail is tabak$@$cims.nyu.edu}\\
        Courant Institute of Mathematical Sciences \\
        New York University}

\maketitle
\begin{abstract}
A novel canonical Hamiltonian formalism is developed for long internal
waves in a rotating environment. This includes the effects of
background vorticity and shear on the waves. By restricting
consideration to flows in hydrostatic balance, superimposed on a
horizontally uniform background of vertical shear and vorticity, a
particularly simple Hamiltonian structure arises, which can be thought
of as describing a nonlinearly coupled infinite collection of shallow
water systems.  The kinetic equation describing the time evolution of
the spectral energy of internal waves is subsequently derived, and a
stationary Kolmogorov solution is found in the high frequency
limit. This is surprisingly close to the Garrett--Munk spectrum of
oceanic internal waves.
\end{abstract}

\begin{centerline}{\Large{\it  submitted to JFM}} \end{centerline}

\section{Introduction} 

The term {\it ocean waves} typically evokes images of surface waves
shaking ships during storms in the open ocean, or breaking
rhythmically near the shore. Yet much of the ocean wave action takes
place underneath the surface, and consists of modulations not of the
air-water interface, but of invisible surfaces of constant
density. These {\it internal waves} are ubiquitous in the ocean,
contain a large amount of energy, and affect significantly the
processes involved in water mixing and transport.

Our knowledge of the typical scales and energy content of oceanic
internal waves advanced significantly through improved and more
widespread observations in the last few decades. In particular, an
empirically based formula that Munk and Garrett developed in the
seventies (now called the Garrett Munk spectra of internal waves)
synthesizes magnificently a seemingly universal distribution of energy
among scales~\cite{GM72,GM75,GM79}. Description of modern
observational work can be found in
\cite{Polzin1,Polzin2,Polzin3,Polzin4,Polzin5}. In particular, in
\cite{Polzin3} the deviation from the Garrett-Munk spectra are
documented, and in \cite{Polzin4} the dissipation rate of turbulent
kinetic energy is measured.  On the theoretical side, this
distribution has been generally understood as due to the effects of
nonlinear interaction among waves, amenable to a description based on
kinetic equations analogous to the ones of statistical mechanics.
Probably the first kinetic equation for internal waves was written in
\cite{Olbers}, though not within the frame of a Hamiltonian formalism.

A comprehensive review of a significant line of work developed in the
seventies and eighties is provided in \cite{Muller} and references
therein. Some important references not cited there are
\cite{Pelinovsky,Voronovich,Miropolskii}. More recent work includes,
for example, \cite{Caillol} where a thorough perturbative
Eulerian--Boussinesq approach in a nonrotating environment is
developed. See \cite{Hines,HinesColin,Chunchuzov} for a detailed
discussion of the relation between the spectral tails of Lagrangian
and Eulerian flow descriptions. One can also use ray theory to study
internal wave scattering, as in \cite{Broutman}, where the
Doppler-spreading of short internal wave packets in the atmosphere and
the ocean is studied or, as in \cite{Eckerman}, where the refraction
of short oceanic internal waves by a spectrum of large amplitude
inertia waves is considered.

Our work differs from the line reviewed in \cite{Muller} in various
ways.  In \cite{Muller} the wave dynamics is formulated in a fully
Lagrangian framework, while our isopycnal formulation is Eulerian in
the horizontal coordinates and Lagrangian in the vertical.  To write
down the equations of motion in the Lagrangian framework, the system's
Lagrangian is expanded in powers of the assumed small displacement of
the fluid parcels. This description is therefore approximate even at
the level of the dynamic equations of motion.  Such a description
fails to adequately describe the advection of small scale waves by
larger scale flows, as well as the interaction of waves with the
vortical part of motion.  This is acknowledged in \cite{Muller}, which
proposes as a challenge the rederivation of the kinetic equation in an
Eulerian framework.  In the present article, we fulfill this program,
and use therefore as a small parameter not the small displacement of
fluid parcels, but the weakness of the nonlinear interaction among
waves.

A fundamental question posed in the eighties is whether the GM spectra
is close to the statistical steady state solution of the kinetic
equation. The discussion in \cite{Muller} indicates that GM is not
inconsistent with the kinetic equation proposed, and may be close to
being a steady state solution.  Recently, the authors have put forward
a study showing that power law spectra which, in the high frequency
limit, are very close to GM, constitute exact analytical steady state
solutions of a kinetic equation for hydrostatic flows described in
isopycnal coordinates \cite{LT}.  The present article extends those
early results, by starting to build a theory that includes frequencies
comparable to the rate of rotation of the Earth, and that accounts for
the existence of large--scale horizontal eddies and vertical shear in
the ocean, over which the internal waves are superimposed.

The main contribution of this paper is the development of a novel
Hamiltonian formalism for the description of internal waves.  Our
approach is general enough to include the effects of the Earth's
rotation, of large--scale eddies and of vertical shear on the waves,
yet exclusive enough in its assumptions to yield a relatively simple,
manageable model. The main assumption is that the waves are long
enough to be in hydrostatic balance, yet they live in horizontal
scales shorter than those characterizing the underlying eddies. This
allows us to consider as unperturbed flow an arbitrary layered
distribution of potential vorticity and vertical shear --that is,
potential vorticity and horizontal velocity profiles which adopt
independent, horizontally uniform values at each depth. Such
hydrostatically balanced, horizontally uniform, vertically varying
profiles are quite representative of long waves in the real ocean;
they arise due to the highly anisotropic nature of the ocean's eddie
diffusivity, which tends to homogenize the flow along isopycnal
surfaces.

Hamiltonian structures for stratified incompressible flows have been a
subject of active research over the last few decades. Though a
complete review of the subject is outside the scope of this paper, we
list here some of the most important results.  The first paper to
derive a Hamiltonian structure for stratified internal waves is
probably \cite{Voronovich}, where a representation is proposed based
on Clebsh-like variables.  The resulting Hamiltonian is an explicit
infinite power series of canonically conjugated variables.  In
\cite{Milder}, a Hamiltonian formalism for internal waves in isopycnal
coordinates is developed.  No hydrostatic approximation is invoked,
and thus the resulting Hamiltonian is expressed as an explicit
power-series in the powers of the assumed small nonlinearity.
Potential problems in using Clebsh variables for stratified flows have
been addressed in \cite{Henyey1,Henyey2}.  Problems of describing the
Hamiltonian structures to describe interaction of wave and vortex mode
is addressed in \cite{Zeitlin1992}. In addition to the references
above, a noncanonical Hamiltonian structure based on a Lie-Poisson
framework has been developed in \cite{HolmHamiltonian}.  More
recently, two broad reviews on Hamiltonian structures for fluids have
been published \cite{Morrison,Salmon}.

We choose to describe the flow in isopycnal coordinates, replacing the
depth $z$ by the density $\rho$ as the independent vertical
coordinate.  The advantage of such semi--Lagrangian description is
manifold.  First, it eliminates the need to handle the vertical
velocity explicitly, which renders the equations much more tractable.
At a deeper level, it greatly simplifies the description of the
interaction between waves and vorticity, since potential vorticity is
preserved along particle trajectories, and these remain on isopycnals
in the absence of mixing. In particular, if the potential vorticity is
uniform throughout an isopycnal surface, it remains so forever. This
is the situations we choose to describe: a profile of vorticity which
varies across surfaces of constant density, but is homogeneous along
them. Such ``pancake--like'' profiles are quite representative of the
intermediate scales in the ocean, smaller than the dominant eddies,
but still containing a significant wealth of internal waves. Modeling
this scenario constitutes an intermediate step between considering
irrotational frameworks, and studying the fully turbulent interaction
between arbitrary profiles of vorticity and waves.

Finally, the equations in isopycnal coordinates are highly reminiscent
of the equations for a shallow single layer of homogeneous fluid. We
exploit this formal analogy to develop a Hamiltonian formalism for
internal waves which extends naturally similar formalisms for shallow
waters.

The plan of this paper is the following. In section
(\ref{sec:Hamiltonians}), we develop a hierarchy of Hamiltonian
descriptions for long waves.  Our goal here is to develop a general
Hamiltonian formalism for nonlinear internal waves in a rotating
environment, superimposed on a background of layered potential
vorticity and horizontally uniform shear. However, we choose to do so
in stages, starting in the simple context of linear, irrotational
shallow water equations, and adding progressively nonlinearity,
stratification, the Coriolis effect, and nontrivial potential
vorticity.  This not only clarifies the logic and essential simplicity
of the structure of the final Hamiltonian, but also yields along the
way a series of Hamiltonian structures for problems of intermediate
simplicity.  In section (\ref{sec:WT}), we derive kinetic equations
for the time evolution of the energy spectrum of internal waves, based
on the Hamiltonian description developed in the previous section. For
this, we constrain ourselves to consider a neutral background with
neither shear nor vorticity.  However, we still include the Coriolis
effect.  Then, in section (\ref{sec:Kol}), we find approximate
stationary solutions to the kinetic equations corresponding to a
direct cascade of energy toward the finest scales, in the
high--frequency limit, where the effects of the Earth's rotation are
negligible. This was the situation considered in \cite{LT}.  Finally,
in section (\ref{sec:concl}), we provide some concluding remarks, and
discuss open problems for further research.

\section{Hamiltonian formalism for long internal waves} 
\label{sec:Hamiltonians}

In this section we develop a Hamiltonian formalism for long internal
waves in a rotating environment. We choose to do so progressing
through a hierarchy of models, which starts with linear, irrotational
shallow water waves in a non--rotating environment, and ends up with
fully nonlinear internal waves in a stratified and rotating
environment, superimposed on an arbitrary layered distribution of
potential vorticity.

\subsection{Linear, Non-Rotating Shallow Waters}

In non--dimensional form, the shallow--water equations take the form
\begin{eqnarray}
  h_t + \nabla \cdot (h \vec u) &=& 0, \label{massnlsw} \\
  {\vec u}_t + (\vec u \cdot \nabla) \vec u + \nabla h &=& 0 \label{momnlsw} \, .
\end{eqnarray}
Here $h$ represents the height of the free--surface, and $\vec{u}$ the
horizontal velocity field. The height $h$ has been normalized by its
mean value $H$, the velocity field $u$ by the characteristic speed $c
= \sqrt{g h}$ (here $g$ is the gravity constant), the horizontal
coordinates by a typical wavelength $L$, and time by $L/c$. Writing
$$ h = 1 + \eta \, , $$
and assuming that $\eta$ and $|u|$ are much smaller than one, one
obtains to leading order the linearized equations
\begin{eqnarray}
  \eta_t + \nabla \cdot \vec u &=& 0 \, \label{masslsw} \\
  \vec u_t + \nabla \eta &=& 0 \, . \label{momlsw}
\end{eqnarray}
At this linear level, the dynamics of waves and vorticity decouple,
with the former satisfying the wave equation, and the latter remaining
constant in time. In particular, if the flow is initially irrotational
(i.e., $\nabla \times \vec u = 0$), it will remain so forever.  Hence
we may restrict our attention here to irrotational flows.  These may
be described by a scalar potential $\phi$, such that
$$
 \vec u = \nabla \phi \, .
$$
For such flows, the system in (\ref{masslsw}, \ref{momlsw})
reduces to
\begin{eqnarray}
  \eta_t + \Delta \phi &=& 0\, , \nonumber \\
  \phi_t + \eta &=& 0. \nonumber
\end{eqnarray}
This system is Hamiltonian, with
\begin{equation}
  \H = \frac{1}{2}
   \int \left(\eta^2 + |\nabla\phi|^2 \right) \, dx .
   \label{Hilsw}
\end{equation}
The Hamiltonian form of the equations is
\begin{eqnarray}
  \eta_t &=& \frac{\delta \H}{\delta \phi} \, ,\label{CE1} \\
  \phi_t &=& - \frac{\delta \H}{\delta \eta} \, . \label{CE2}
\end{eqnarray}

Notice that the Hamiltonian in (\ref{Hilsw}) is the sum of the
potential and kinetic energy of the system. The former is actually
given by $\frac{1}{2} (1+\eta)^2$, but the difference can be absorbed
by a gauge transformation of the potential $\phi$. Our goal is to
preserve the essential simplicity of this formulation when we add
nonlinearity, ambient rotation and stratification.

\subsection{Nonlinear, Non-Rotating Shallow Waters}

For the fully nonlinear shallow--water equations in (\ref{massnlsw},
\ref{momnlsw}), waves and vorticity no longer decouple (in fact, the
nonlinear interaction of waves and vorticity is among the main
theoretical obstacles to a full description of turbulence.)  However,
it is still true that a flow which starts irrotational stays so
forever. Hence we may restrict ourselves to considering this scenario,
introduce again the scalar potential $\phi$, and rewrite
(\ref{massnlsw}, \ref{momnlsw}) in the form
\begin{eqnarray}
  h_t + \nabla \cdot (h \nabla \phi) &=& 0 \, ,\nonumber\\ \phi_t +
  \frac{1}{2} |\phi|^2 + h &=& 0 \, .\nonumber\\
  \label{NonlinearShallowWater}
\end{eqnarray}
This system is also Hamiltonian, with
\begin{equation}
  \H =  \frac{1}{2}\int
   \left(  h^2 +  h |\nabla\phi|^2 \right) \, 
   dx \, ,
\label{Hinsw}
\end{equation}
and canonical equations
\begin{eqnarray}
   h_t &=& \frac{\delta \H}{\delta \phi} \, , \nonumber \\
  \phi_t &=& - \frac{\delta \H}{\delta h} \nonumber \, .
\end{eqnarray}

In this case, the Hamiltonian is the sum of the potential and kinetic
energy, without qualifications.

\subsection{Nonlinear, Non-Rotating, Internal Waves} 

The non--dimensional equations of motion for long internal waves in an
incompressible, stratified fluid with hydrostatic balance, are given
by
\begin{eqnarray}
\frac{d {\vec u}}{d t} + \frac{{\bf \nabla} P}{\rho} = 0,& \ \ \ &
P_z+\rho = 0, \nonumber\\
\frac{ d \rho}{d t} =0,& \ \ \ &
{\bf \nabla}\cdot {\vec u} + w_z = 0 \, , \nonumber
\end{eqnarray}
where ${\vec u}$ and $w$ are the horizontal and vertical components of
the velocity respectively, $P$ is the pressure, $\rho$ the density,
$\nabla = (\partial_x, \partial_y)$ the horizontal gradient operator,
and
$$ 
  \frac{d}{d t} = \frac{\partial}{\partial t} +
    {\vec u}\cdot{\bf \nabla} + w \frac{\partial}{\partial z}
$$
is the Lagrangian derivative following a particle.  

Changing to isopycnal coordinates ($x,y,\rho,t$) , where the roles of
the vertical coordinate $z$ and the density $\rho$ as independent and
dependent variables are reversed, the equations become:
\begin{eqnarray}
  \frac{D \vec{u}}{D t} + \frac{\nabla M}{\rho} &=& 0\, ,\nonumber  \\
  M_{\rho} &=& z \, , \nonumber \\
  z_{\rho,t} + \nabla \cdot \left(z_{\rho} \vec{u} \right)
	 &=& 0 \, . \nonumber\\ \label{IsopycnalInternalWaveEqn}
\end{eqnarray}
Here $\vec{u} = (u,v)$ is the horizontal component of the velocity
field, $\nabla = (\partial_x, \partial_y)$ is the gradient operator
along isopycnals, $\frac{D }{D t} = \partial_t + \vec{u} \cdot
\nabla$, and $M$ is the Montgomery potential~\cite{Kushmar},
 $$M=P+\rho\,z.$$ For flows which are irrotational along isopycnal
surfaces, we introduce the velocity potential
$$\vec u=\nabla \phi.$$ Such a
substitution allows us to integrate (\ref{IsopycnalInternalWaveEqn}) once
and illuminate $z$, after which these equations reduce to the pair
\begin{eqnarray}
  \phi_t + \frac{1}{2} |\nabla \phi|^2 + \frac{1}{\rho} 
     \int^{\rho}\int^{\rho_2} \frac{\Pi}{\rho_1} \, d\rho_1 \, d\rho_2 &=& 0 
\, ,\nonumber
\\
   \Pi_t + \nabla \cdot \left(\Pi\, \nabla \phi \right)
    &=& 0 \, ,\nonumber\\ \label{PRLeqOfmotion}
\end{eqnarray}
where we have introduced the variable
$$\Pi = \rho M_{\rho \rho} = \rho z_{\rho}.$$ 

This variable $\Pi$ has at least two physical interpretations.  One is
that of density in isopycnal coordinates, since

$$ \Pi \, d\rho = \rho \, dz \, . $$
The other is that of a measure of the stratification, namely the
relative distance between neighboring isopycnal surfaces, since this
distance $dz$ is given by
$$ dz = \Pi \frac{d\rho}{\rho} \, . $$ 

Notice the similarity between (\ref{PRLeqOfmotion}) and the equations
(\ref{NonlinearShallowWater}) for nonlinear shallow water.  Internal
wave equations could be viewed as a system of infinitely many, coupled
shallow water equations. This analogy allows us to identify a natural
Hamiltonian structure for internal waves.

 The variable $\Pi$ is also the canonical conjugate of $\phi$, 
\begin{equation}
\Pi_t=\frac{\delta {\H}}{\delta \phi}, \ \ \ \ \ \ \
\phi_t=-\frac{\delta {\H}}{\delta \Pi} \, ,
\label{Canonical}
\end{equation}
under the Hamiltonian flow given by
\begin{equation}
  \H = \frac{1}{2}\int \left( \Pi \, |\nabla \phi|^2 -
    \left|\int^{\rho} \frac{\Pi}{\rho_1} \, d\rho_1\right|^2 \right) d {\bf r} d \rho\, .
\end{equation}
The first term in this Hamiltonian clearly corresponds to the kinetic
energy of the flow; that the second term is in fact the potential
energy follows from the simple calculation
\begin{eqnarray*}
  \frac{1}{2} \left|\int^{\rho} \frac{\Pi}{\rho_1} \, d\rho_1\right|^2
    d\rho = \frac{1}{2} \left|\int^{z} dz\right|^2 d\rho =&& \\ =
    \frac{1}{2} z^2 d\rho = - \rho \, z \, dz + d \left(\frac{1}{2}
    \rho\, z^2 \right) && \, ,
\end{eqnarray*}
so
\begin{equation}
  -\int_{\rho(z_t)}^{\rho(z_b)}
  \frac{1}{2} \left|\int^{\rho} \frac{\Pi}{\rho_1} \,
  d\rho_1\right|^2
  d\rho = \int_{z_b}^{z_t} \rho \, z \, dz -
  \frac{1}{2} \rho\, z^2 \Bigg|_{z_b}^{z_t}\, , \nonumber
\end{equation}
where $b$ and $t$ stand for bottom and top respectively, and the
boundary conditions are usually such that the integrated term at the
end is a constant.

\subsection{Linear Shallow Waters in a Rotating Environment}

In a rotating environment, the linearized shallow--water equations are
\begin{eqnarray}
  \eta_t + \nabla \cdot \vec{u} &=& 0\, ,\nonumber \\
  \vec{u}_t + \nabla \eta + \vec{u}^{\perp} &=& 0\, .\nonumber
\end{eqnarray}
Here
$$ \vec{u} = \left( \begin{array}{r}
  u \\
  v
\end{array} \right) \, $$
is the velocity field, and
$$ \vec{u}^{\perp} = \left( \begin{array}{r}
  -v \\
   u
\end{array} \right) \, . $$
The Coriolis parameter $ f $ has been absorbed in the
nondimensionalization of time, so it is effectively equal to one.

These equations do not preserve vorticity, so irrotationality
cannot be assumed. However, they preserve the {\it potential
vorticity}
\begin{equation}
  q = v_x - u_y - \eta \, .
\end{equation}
The assumption corresponding to irrotationality in the non-rotating
case is therefore that of zero potential vorticity, i.e. $q=0$.  We
can in fact generalize this hypothesis, and consider an arbitrary,
though constant, potential vorticity. We shall employ such
generalization when we consider internal waves in a rotating
environment.  In order to exploit this assumption, it is convenient to
decompose the flow into a potential and a divergence-free part:
\begin{equation}
  \vec{u} = \nabla \phi + \nabla^{\perp} \psi \, ,
\label{decomp}
\end{equation}
where 
\begin{equation}
  \nabla^{\perp} =\left( \begin{array}{r}
	-\partial_y \\ 
	 \partial_x 
	 \end{array} \right) \, .
\end{equation} 
In terms of $\phi$ and $\psi$, the equations take the form
\begin{eqnarray}
  \eta_t + \Delta \phi &=& 0\nonumber \, , \\
  \phi_t + \eta - \psi &=& 0 \nonumber \, ,\\
  \psi_t + \phi &=& 0 \, .
\end{eqnarray}
The condition of zero potential vorticity takes the form
\begin{equation}
  q = v_x - u_y - \eta = \Delta \psi - \eta = 0 \, ,
\end{equation}
so the system above reduces to
\begin{eqnarray}
  \eta_t + \Delta \phi &=& 0 \\
  \phi_t + \eta - \Delta^{-1} \eta &=& 0 \, .
\end{eqnarray}  
This system is Hamiltonian, with canonical variables $\phi$ and
$\eta$, and Hamiltonian
\begin{equation}
  \H = \frac{1}{2} \int \left( \left|\nabla\phi + 
\nabla \Delta^{-1} \eta \right|^2 +
	   \eta^2 \right) d  x \, .
\end{equation}
Again, the Hamiltonian agrees with the total energy of the system.

\subsection{Rotating Nonlinear Shallow Waters}

The fully nonlinear equations for shallow waters in a rotating
environment are
\begin{eqnarray}
  h_t + \nabla \cdot (h \vec u) &=& 0 ,\label{hequation}\\
  \vec{u}_t + (\vec{u} \cdot \nabla) \vec{u} + \nabla h + \vec{u}^{\perp} &=& 0 \,\label{uequation} .
\end{eqnarray}

The statement of conservation of potential vorticity now takes the
form~(Sec 12-2 in \cite{Kushmar})
\begin{equation}
  \frac{D}{Dt} \left(\frac{1+v_x-u_y}{h}\right) = 0 \, 
\end{equation}
(That is: the total vorticity of a vertical column of water divided by
its height remains constant as the column moves.)  The unperturbed
state has
$$q=\frac{1+v_x-u_y}{h}=q_0,$$ where $q_0$ is an arbitrary
 (prescribed) potential vorticity, so this is the hypothesis to make
 for the analogue of irrotational flows:
\begin{equation}
  q_0 h = 1 + v_x - u_y \, .
  \label{nlpotvort}
\end{equation}
We introduce the potentials $\phi$ and $\psi$ as in (\ref{decomp}),
and use the fact that
$$(\vec u \cdot \nabla )\vec u = \frac{1}{2}\nabla | \nabla \phi +
\nabla^{\perp}\psi|^2 + \Delta \psi \left(\nabla^\perp \phi - \nabla
\psi \right),$$ to rewrite (\ref{uequation}) as
$$\nabla \phi_t + \nabla^\perp \psi_t + \frac{1}{2}\nabla | \nabla
\phi + \nabla^{\perp}\psi|^2 +\Delta \psi \left(\nabla^\perp \phi -
\nabla \psi \right)+\nabla h +\nabla^\perp\phi-\nabla\psi = 0.$$
Taking the divergence and the two--dimensional curl $\nabla^{\perp} \cdot$
 of the above
equations, we
obtain the following pair:
\begin{eqnarray}
    \phi_t + \frac{1}{2} |\nabla\phi+\nabla^{{\perp}}\psi|^2  +
  \Delta^{-1} \nabla \cdot 
    \left[\Delta\psi (\nabla^{{\perp}}\phi - \nabla\psi) \right]
  + h -\psi&=& 0 , \nonumber \\
  \psi_t + 
  \Delta^{-1} \nabla^{{\perp}} \cdot 
      \left[\Delta\psi (\nabla^{{\perp}}\phi - \nabla\psi) \right]+\phi
      &=& 0 \, .\nonumber
\end{eqnarray}

By noticing that
\begin{eqnarray}
-\psi= \Delta^{-1} \nabla \cdot (\nabla^{{\perp}}\phi - \nabla\psi),
\nonumber\\ \phi= \Delta^{-1} \nabla^\perp \cdot (\nabla^{{\perp}}\phi
- \nabla\psi),\nonumber
\end{eqnarray}
we can rewrite these equations, together with
(\ref{hequation}) in the form
\begin{eqnarray}
  h_t + \nabla \cdot (h\, (\nabla \phi + \nabla^{{\perp}} \psi)) &=& 0 ,
\nonumber  \\
  \phi_t + \frac{1}{2} |\nabla\phi+\nabla^{{\perp}}\psi|^2  +
  \Delta^{-1} \nabla \cdot 
    \left[(1+\Delta\psi) (\nabla^{{\perp}}\phi - \nabla\psi) \right]
  + h &=& 0 , \nonumber \\
  \psi_t + 
  \Delta^{-1} \nabla^{{\perp}} \cdot 
      \left[(1+\Delta\psi) (\nabla^{{\perp}}\phi - \nabla\psi) \right]
      &=& 0 \, .\nonumber
\end{eqnarray}

The constraint (\ref{nlpotvort}) on the potential vorticity takes
the form
\begin{equation}
  1 + \Delta \psi = q_0 h \, ,
\end{equation}
under which the equations above reduce to the pair
\begin{eqnarray}
  h_t + \nabla \cdot (h\, (\nabla \phi + \nabla^{{\perp}}
    \Delta^{-1}(q_0 h-1))) &=& 0 \nonumber \\ \phi_t + \frac{1}{2}
    (\nabla\phi+\nabla^{{\perp}}\Delta^{-1}(q_0 h-1))^2 + && \nonumber \\
    q_0 \Delta^{-1} \nabla \cdot \left[h\, (\nabla^{{\perp}}\phi -
    \nabla\Delta^{-1}(q_0 h-1)) \right] + h &=& 0 \, .\nonumber
\label{NonlinearShallowWaterCompact}
\end{eqnarray}
These equations are Hamiltonian, with conjugate variables $\phi$
and $h$, and Hamiltonian
\begin{equation}
  \H =\frac{1}{2} \int\left(
 h\, \left|\nabla\phi+\nabla^{{\perp}}\Delta^{-1}(q_0 h-1)\right|^2
	  +  h^2 \right) d {\bf r} \, ,
\end{equation}
representing again the sum of kinetic and potential energies.

\subsection{Nonlinear Internal Waves in a Rotating Environment}

The equations for long internal waves in a rotating environment are
particularly simple when written in the isopycnal coordinates
$(x,y,\rho,t)$; they take the form in (\ref{PRLeqOfmotion}) with an
extra term $ \vec{u}^{\perp}$ due to the Coriolis force:
\begin{eqnarray}
  \frac{D \vec{u}}{D t} + \vec{u}^{\perp} + 
    \nabla  \frac{1}{\rho}
  \int^{\rho}\int^{\rho_2} \frac{\Pi-\Pi_0}{\rho_1} \, d\rho_1 \, d\rho_2 &=& 0,
  \nonumber   \\
    \Pi_t + \nabla \cdot \left(\Pi\, \vec{u} \right)
    &=& 0 \, .\nonumber\\
    \label{LIW}
\end{eqnarray}
where
$$ \Pi_0 = \Pi_0(\rho) $$
is a reference stratification profile, that we introduce here for
future convenience.

The expression for the potential vorticity in these coordinates is
\begin{equation}
 q = \frac{1+v_x-u_y}{\Pi} \, ,
\end{equation}
and it satisfies
\begin{equation}
  \frac{Dq}{Dt} = 0 \, \label{Vorticity}.
\end{equation}
Notice that the advection of potential vorticity in (\ref{Vorticity})
takes place exclusively along isopycnal surfaces. Therefore, an
initial distribution of potential vorticity which is constant on
isopycnals, though varying across them, will never change.  Hence we
shall propose that
\begin{equation}
  q = q_0(\rho) \, ,
  \label{PV}
\end{equation}
where $q_0(\rho)$ is an arbitrary function; i.e., one may assign any
constant potential vorticity to each isopycnal surface. This is a
highly nontrivial extension of the irrotational waves of the previous
sections. Extending our description further to include general
distributions of potential vorticity, varying even within surfaces of
constant density, would necessarily complicate its Hamiltonian
formulation, making it loose its natural simplicity. In fact, the
problem of interaction between vorticity and waves is that of fully
developed turbulence, which escapes the scope of our description.
However, the ``pancake--like'' distributions of potential vorticity
that we propose are common in stratified fluids, particularly the
ocean and the atmosphere. They arise due to the sharp contrast between
the magnitudes of the turbulent diffusion along and across
isopycnals. Thus potential vorticity is much more rapidly homogenized
along isopycnals than throughout the fluid, yielding the ``pancakes''.
As we show below, even waves super--imposed on such a general and
realistic distribution of potential vorticity admit a rather simple
Hamiltonian description.

In order to isolate the wave dynamics satisfying the constraint
(\ref{PV}), we decompose the flow into a potential and a
divergence-free part as in (\ref{decomp}).  In terms of the potentials
$\phi$ and $\psi$, (\ref{PV}) and (\ref{Vorticity}) yield
\begin{equation}
  1 + \Delta \psi = q_0 \Pi \, ,\nonumber
\end{equation}
and, repeating the same steps as in nonlinear rotating shallow waters,
the equations in (\ref{LIW}) reduce to the pair
\begin{eqnarray}  
 \Pi_t + \nabla \cdot \left(\Pi\, \left(\nabla \phi + \nabla^{{\perp}} \Delta^{-1}
 \left(q_0 \Pi-1\right)\right)\right)
  &=& 0 \, , \nonumber\\
  \phi_t + \frac{1}{2} 
  \left|\nabla\phi+ \nabla^{{\perp}}\Delta^{-1} \left(q_0 \Pi-1\right)
  \right|^2 
  && \nonumber \\ +
   \Delta^{-1} \nabla \cdot
   \left[ q_0 \Pi \,
    \left(\nabla^{{\perp}}\phi - 
    \nabla\Delta^{-1}\left(q_0 \Pi-1\right)\right) \right] 
    && \nonumber \\
    + \frac{1}{\rho} 
    \int^{\rho}\int^{\rho_2} \frac{\Pi-\Pi_0}{\rho_1} \, d\rho_1 \,
    d\rho_2  &=& 0 \, .\nonumber\\ \label{LIWHAM}
\end{eqnarray}
This pair is Hamiltonian, with conjugated variables $\phi$ and $\Pi$,
i.e. it can be written as
\begin{equation}
\Pi_t=\frac{\delta {\H}}{\delta \phi},
\ \ \ \  \ \ \
\phi_t=-\frac{\delta {\H}}{\delta \Pi} \, .\nonumber
\end{equation}
where the  Hamiltonian is given by
\begin{equation}
  \H = \int \left[ \frac{1}{2} \Pi \, 
  \left|\nabla \phi+  \nabla^{{\perp}}\Delta^{-1}\left(q_0 \Pi-1\right)
  \right|^2 -
  \frac{1}{2} \left|\int^{\rho} \frac{\Pi-\Pi_0}{\rho_1} \, d\rho_1\right|^2 
  \right] d\rho d\vec{r} \, .
  \label{HT}
\end{equation}
Again, this Hamiltonian represents the sum of the kinetic and
potential energy of the flow.

Notice the similarity of our description of internal waves with the
Hamiltonian formulation for free--surface waves introduced in
\cite{Z68a,MilesHamiltonian}.  There, it was shown that the
free--surface displacement and the three--dimensional velocity
potential evaluated at the free surface are canonical conjugate
variables. In our case, the canonical conjugate variables are also a
displacement and a velocity potential, though the velocity potential
in (\ref{HT}) is for the two--dimensional flow along isopycnal
surfaces, and the displacement is the relative distance between
neighboring isopycnal surfaces, as described above.

Looking back, we could have included some vorticity from early on;
there was no need to take it equal to zero, as the last section shows.
In shallow waters, it could have been any constant; in internal waves,
any function of the density. It is clear though that, if one wanted to
include arbitrary vorticity distributions, one would need to go fully
Lagrangian, to exploit the fact that vorticity is preserved along
particle paths.  This would make the Hamiltonian structure less
appealingly simple.

The key steps taken here for finding a simple Hamiltonian structure
for internal waves, could be summarized as follows:

\begin{enumerate}

  \item To consider long waves in hydrostatic balance. This, together
        with the choice of isopycnal coordinates, leads to a system of
        equations formally equivalent to an infinite collection of
        coupled shallow--water systems. This analogy allows us to
        generalize the relatively simple Hamiltonian structure of
        irrotational shallow--waters to the richer domain of internal
        waves.

  \item To decouple waves from vorticity, by assuming the latter to be
        either zero, constant or uniform along isopycnal surfaces,
        with an arbitrary dependence on depth. This is facilitated by
        the choice of a flow description in isopycnal coordinates.

  \item To realize that the potential $\phi$ is a good candidate
        canonical variable, and that its conjugate is the height $h$
        for shallow waters, and the surrogate $\Pi$ for density in the
        isopycnal formulation of internal waves.

  \item To introduce nonlocal operators into the Hamiltonian.  These
        arise naturally from the ``elliptic'' constraints of
        hydrostatic balance and layered potential vorticity. Despite
        its unusual look, the Hamiltonian is invariably just the sum
        of the standard kinetic and potential energies, integrated
        over the domain.

\end{enumerate}
The assumptions of hydrostatic balance and horizontally uniform
background vorticity and shear, which simplify notoriously the
description of the flows, are quite realistic for a wide range of
ocean waves.

\section{Weak turbulence theory}
\label{sec:WT}

In this section, we apply the formalism of wave turbulence theory to
derive a kinetic equation, describing the time evolution of the energy
spectrum of internal waves. In order to do this, we need to assume
that the waves are weakly nonlinear perturbations of a background
state. In principle, we could adopt for this state an arbitrary
background distribution of (layered) potential vorticity, vertical
shear and stratification. To make the derivation clear, however, we
focus here on the case with zero shear and zero potential vorticity,
and a stratification profile with constant buoyancy frequency. Even
though the mechanics for deriving the kinetic equation in the more
general setting are entirely similar (though more cumbersome), the
tools at our disposal for finding relevant exact solutions to these
equations, which we actually use in section \ref{sec:Kol}, are only
applicable to the case with the simplest background.

To leading order in the perturbation, we obtain linear waves, with
amplitudes modulated by the nonlinear interactions. These linear waves
have, in general, a complex vertical structure (they are
eigenfunctions of a differential eigenvalue problem), but reduce, in
our case, to sines and cosines~\cite{Gill}.

Let us now take (\ref{HT})  and rewrite it in dimensional form:
\begin{equation}
  \H = \int \left[ \frac{1}{2} \Pi \, 
  \left|\nabla \phi+  \nabla^{{\perp}}\Delta^{-1}\left(q_0 \Pi-f\right)
  \right|^2 -
  \frac{g}{2} \left|\int^{\rho} \frac{\Pi-\Pi_0}{\rho_1} \, d\rho_1\right|^2 
  \right] \, .
  \label{HTL}
\end{equation}

Here $ f $ is the Coriolis parameter, $g$ is the acceleration due to
gravity.  Note that $$[\Pi]={\rm Length}, \ \ \ \ [\phi]=\frac{\rm
Length^2}{Time}.$$

The potential vorticity is, in dimensional form,
$$ q = \frac{f + v_x - u_y}{\Pi}.  $$
In the calculations that follow, we shall consider flows which are
perturbations of a state at rest, stratified but without
vorticity. When this is the case, $v_x - u_y$ is zero to leading
order, and we have the following relation between the potential
vorticity profile $q_0$ and the stratification profile $\Pi_0$:
\begin{equation}
  q_0(\rho) = \frac{f}{\Pi_0(\rho)} \, .
  \label{Pi0}
\end{equation}
Moreover, the definition of $\Pi$ implies that
\begin{equation}
  \Pi_0 = -\frac{g}{N^2} \, ,
  \label{buoyancy}
\end{equation}
where $N(\rho)$ is the buoyancy frequency, which we shall consider
here to be a constant.

For the subsequent calculations it will be convenient to decompose the
potential $\Pi$ into its equilibrium value and its deviation from
it. Therefore let us redefine $$\Pi \to \Pi_0+\Pi \, . $$ Then the
Hamiltonian takes the following form:
{\begin{equation}
  \H = \int d {\vec r} d \rho \left[ \frac{1}{2} \left( -\frac{g}{N^2}+\Pi \right) \, 
  \left|\nabla \phi- \frac{f N^2 }{g} \nabla^{{\perp}}\Delta^{-1} \Pi
  \right|^2 -
  \frac{g}{2} \left|\int^{\rho} \frac{\Pi}{\rho_1} \, d\rho_1\right|^2 
  \right] \, .
  \label{HTL2}
\end{equation}

It can be represented as a sum of a quadratic and a cubic part:
\begin{eqnarray}
\H = \H_{\rm linear} +\H_{\rm nonlinear},\nonumber\\ \H_{\rm linear}=
\int  d {\vec r} d \rho
\left[ -\frac{g}{2 N^2} \left|\nabla \phi-\frac{N^2
 f }{g}\nabla^{{\perp}}\Delta^{-1} \Pi \right|^2 - \frac{g}{2}
\left|\int^{\rho} \frac{\Pi}{\rho_1} \, d\rho_1\right|^2 \right] \, ,
\nonumber\\ \H_{\rm nonlinear} = \frac{1}{2} \int  d {\vec r} d \rho
\Pi \left|\nabla
\phi-\frac{N^2  f }{g}\nabla^{{\perp}}\Delta^{-1} \Pi \right|^2
\label{H2}
\end{eqnarray}

Let us use the Fourier transformation:
\begin{eqnarray}
\Pi({\vec{r}},\rho) = \frac{1}{(2 \pi)^{3/2}}\int \Pi_{\vec p} e^{ i
{\vec R} {\vec p}} d {\vec p}, \nonumber\\ \phi({ \vec r},\rho) =
\frac{1}{(2 \pi)^{3/2}}\int 
\phi_{\vec p} e^{ i {\vec R} {\vec p}} d
{\vec p}, \nonumber\\
{\vec p}=
( {\vec{k}},m), \ \ 
{\vec R}=( {\vec{r}},\rho) . \nonumber
\end{eqnarray}
Note that the operator $\nabla^{{\perp}}\Delta^{-1}$ has a simple
representation in Fourier space:
$$\nabla^{{\perp}}\Delta^{-1} \Pi({\vec R}) =
-\frac{i}{(2\pi)^{3/2}}\int d {\vec p} \frac{ {\vec k^{{\perp}}}}{k^2}
e^{i {\vec p}{\vec R} }\Pi_{\vec p}, \ \ \ {\vec k^{{\perp}}} =
(-k_y,k_x) \ .$$ Since in the ocean, $\rho$ deviates from its
equilibrium value $\rho_0$ by no more then 3\%, it is natural to make
the Boussinesq approximation, replacing the density by a reference
value $\rho_0$:
$$\frac{g}{2}\left| \int^{\rho} \frac{\Pi}{\rho_1}\, d \rho
\right|\simeq \frac{g}{2\rho_0}\left|\int^{\rho} \Pi d \rho \right|.$$
Then
\begin{eqnarray}
\H_{\rm linear} = -\frac{1}{2}\int d {\vec p} \left( \frac{g}{N^2} k^2
|\phi_{\vec p}|^2 + \left(\frac{N^2 f ^2}{g k^2}+ \frac{g}{\rho_0^2
  m^2} \right)|\Pi_{\vec p}|^2 \right)\ ,\nonumber\\ \H_{\rm
  nonlinear} = \frac{1}{2}\int d \p_1 d \p_2 d \p_3
\delta(\p_1+\p_2+\p_3) \left( -\k_2\cdot\k_3 \Pi_{\p_1} \phi_{\p_2}
\phi_{\p_3} - \frac{N^4 f ^2}{g^2} \frac{\k_2\cdot\k_3}{k_2^2
  k_3^2}\Pi_{\p_1}\Pi_{\p_2}\Pi_{\p_3} - \nonumber \right. \\ \left.
2 \frac{N^2 f
}{g}\frac{\k_2\cdot\k_3^{{\perp}}}{k_3^2}\Pi_{\p_1}\Phi_{\p_2}\Pi_{\p_3}\right)\,
.\nonumber\\
\label{WT1}
\end{eqnarray}
From now on it will be convenient to use the following short--hand notation:
\begin{enumerate}
\item
$\int d123$ instead of $d {\p_1} d {\p_2} d {\p_3}$, 
\item $\delta_{1+2+3}$ instead of $\delta(\p_1+\p_2+\p_3)$,
\item $\Pi_{i}$ and $\Phi_i$ instead of $\Pi_{{\bf p_i}}$ and
$\Phi_{{\bf p_i}}$.
\end{enumerate}
Then the last formula can be written in a more compact form:
\begin{eqnarray}
\H_{\rm nonlinear} = \frac{1}{2}\int d123 \delta_{1+2+3} \left(
-\k_2\cdot\k_3 \Pi_1 \phi_2 \phi_3 - \frac{N^4 f ^2}{g^2}
\frac{\k_2\cdot\k_3}{k_2^2 k_3^2}\Pi_1\Pi_2\Pi_3 - \nonumber
\right. \\ \left.  2 \frac{N^2 f
}{g}\frac{\k_2\cdot\k_3^{{\perp}}}{k_3^2}\Pi_1\Phi_2\Pi_3\right)\nonumber\\
\label{WT2}
\end{eqnarray}

The canonical equations of motions (\ref{Canonical}) form a pair of
real equations. Their Fourier transformation gives a pair of two
complex equations, yet not independent. To reduce this pair to one
complex equation, one performs the transformation
\begin{eqnarray}
\phi_\p=\frac{i}{\sqrt{2}\sqrt{f_\p}}
\left(a_\p-
a^*_{-\p}\right)\, ,\nonumber\\
\Pi_\p =\frac{\sqrt{f_\p}}{\sqrt{2}}\left(a_\p+a^*_{-\p}\right)\, . \nonumber\\
\label{transformationToSingleEquation}
\end{eqnarray}
Here $f_\p$ is a real, positive and even, otherwise arbitrary
function.

This transformation turns the pair of canonical equation of motion
(\ref{Canonical}) into a single equation for the complex variable
$a_\p$:
\begin{equation}
i\frac{\partial}{\partial t} a_{\bf p} =
  \frac{\partial {\H}}{\partial a_{\bf p}^*} \, .
\end{equation}
The following choice of $f_\p$
$$f_\p = \sqrt{\frac{ g k^2 }{N^2} \left(\frac{N^2 f ^2}{g k^2} +
\frac{g}{\rho_0^2 m^2}\right) ^{-1} }$$ diagonalizes the quadratic
part of a Hamiltonian, bringing it to the following form:
$${\H_{linear}}=\int \omega_\p \, |a_\p|^2 \, d \p,$$
where $\omega_\p$ is the dispersion relation for linear
internal waves in isopycnal coordinates:
\begin{equation}
\omega_\p=\sqrt{ f ^2+\frac{g^2 k^2 }{\rho_0^2 m^2 N^2}}.
\label{InternalWavesDispersion}
\end{equation} 
(In the more familiar Eulerian framework, the dispersion relation
transforms into
$$ 
\omega_\p=\sqrt{ f ^2+\frac{N^2 k^2 }{{m_*}^2}},
$$
where $m_*$, the vertical wavenumber in $z$ coordinates, is given by
$m_* = -\frac{g}{\rho_0 N^2} m$ .)

With such a choice of $f_{p}$ the transformations
(\ref{transformationToSingleEquation}) take the following form:
\begin{eqnarray}
\phi_\p=\frac{i N \sqrt{\omega_\p}}{\sqrt{2 g}k}
\left(a_\p-
a^*_{-\p}\right)\, ,\nonumber\\
\Pi_\p =\frac{\sqrt{g} k}{\sqrt{2\omega_\p}N}\left(a_\p+a^*_{-\p}\right)\, . \nonumber\\
\label{transformationToSingleEquation2}
\end{eqnarray}
In terms of $a_\k$, the Hamiltonian (\ref{WT1}) reads 
\begin{eqnarray}
{ \H}=\int \omega_p \, |a_{\p}|^2 \, d {\p} + \nonumber \\
\int
 V_{ {\p_1} {\p_2} {\p_3}} \left( a_{\p_1}^* a_{\bf
p_2}^* a_{\p_3} + a_{\p_1} a_{\p_2}^* a_{\p_3}^*\right)\,
\delta_{ {\p_1} -{\p_2} - {\p_3}} \, d{\p}_{123} +\nonumber \\
\int 
 U_{ {\p_1} {\p_2} {\p_3}} \left( a_{\p_1}^* a_{\bf
p_2}^* a_{\p_3}^* + a_{\p_1} a_{\p_2} a_{\p_3}\right) \,
\delta_{ {\p_1} + {\p_2} +{\p_3}}\, d {\p}_{123}  .\nonumber\\
\label{HAM}
\end{eqnarray}
This is a standard three-wave Hamiltonian of wave turbulence theory.
The calculation of the interaction coefficients is a straightforward
task, yielding
\begin{eqnarray}
V^1_{23}&=& I^1_{23}+J^1_{23}+K^1_{23},\nonumber\\ 
I^1_{23}&=&-\frac{N}{4\sqrt{2 g}}
\left(
\frac{\k_2\cdot\k_3}{k_2 k_3}\sqrt{\frac{\omega_2 \omega_3}{\omega_1}} k_1
+
\frac{\k_1\cdot\k_3}{k_1 k_3}\sqrt{\frac{\omega_1 \omega_3}{\omega_2}} k_2
+
\frac{\k_1\cdot\k_2}{k_1 k_2}\sqrt{\frac{\omega_1 \omega_2}{\omega_3}} k_3
\right)\, ,\nonumber\\
J^1_{23}&=&\frac{N f ^2}{4\sqrt{ 2 g \ \omega_1 \omega_2 \omega_3} }\left(
\frac{\k_2\cdot\k_3}{k_2 k_3}{k_1}
-\frac{\k_1\cdot\k_3}{k_1 k_3}{k_2}
-\frac{\k_1\cdot\k_2}{k_1 k_2}{k_3}
\right)\, ,
\nonumber \\
K^1_{23} &=& \frac{i  f  N}{\sqrt{2 g}}\frac{\k_2\cdot\k_3^{{\perp}}}{k_1 k_2 k_3}
\left(\sqrt{\frac{\omega_2}{\omega_1\omega_3}}(k_1^2-k_3^2) + 
      \sqrt{\frac{\omega_1}{\omega_2\omega_3}}(k_2^2-k_3^2) +
      \sqrt{\frac{\omega_3}{\omega_1\omega_2}}(k_2^2-k_1^2) 
\right) \, ,
\nonumber \\ \label{MatrixElement}
\end{eqnarray}
where we have used the fact that $\vec{k_1} = \vec{k_2} + \vec{k_3}$. 

Following wave turbulence theory, one proposes a perturbation
expansion in the amplitude of the nonlinearity. This expansion gives
to leading order, linear waves. Then one allows the amplitude of the
waves to be slowly modulated by resonant nonlinear interactions. This
modulation is described by an approximate {\it kinetic
equation}~\cite{ZLF} for the ``number of waves'' or wave-action
$n_{\bf p}$, defined by
$$n_{\bf p} \delta({\bf p} - {\bf p'}) = \langle a_{\bf p}^* a_{\bf
p'}\rangle \, .$$
This kinetic equation is the classical analog of the Boltzmann
collision integral.  The basic ideas for writing down the kinetic
equation to describe how weakly interacting waves share their energies
go back to Peierls.  The modern theory has its origin in the works of
Hasselmann \cite{Hass,Hass1}, Benney and Saffmann \cite{Ben},
Kadomtsev \cite{K}, Zakharov \cite{Z68a,Z68b,ZLF}, and Benney and
Newell \cite{Newell,Newell1}. The derivation of kinetic equations
using the wave turbulence formalism can be found, for instance, in
\cite{ZLF,majda}.
For the three-wave Hamiltonian (\ref{HAM}), the kinetic equation
reads:
\begin{eqnarray}
\frac{d n_{\bf p}}{dt} = \pi \int
 |V_{p p_1 p_2}|^2 \, f_{p12} \,
\delta_{{{\bf p} - \bf{p_1}-\bf{p_2}}} \, \delta_{\omega_{{\bf p}}
-\omega_{{\bf{p_1}}}-\omega_{{\bf{p_2}}}}
d {\bf p}_{12} \, ,
\nonumber \\
-2\pi\int
 \, |V_{p_1 p p_2}|^2\, f_{1p2}\, \delta_{{{\bf p_1} - \bf{p}-\bf{p_2}}} \,
  \delta_{{\omega_{{\bf p_1}} -\omega_{{\bf{p}}}-\omega_{{\bf{p_2}}}}}\Big)
\, d {\bf p}_{12} \, ,
\label{KEinternal}
\end{eqnarray}
where $ f_{p12} = n_{{\bf p_1}}n_{{\bf p_2}} -
n_{{\bf p}}(n_{{\bf p_1}}+n_{{\bf p_2}}) \, .
$

Assuming horizontal isotropy, one can average (\ref{KEinternal}) over
all horizontal angles, obtaining
\begin{eqnarray} \label{KEinternalAveragedAngles}
 \frac{d n_p}{d t}
= \frac{1}{k}\int
\left(R^k_{12} - R^1_{k2} - R^2_{1k} \right) \,
d k_1 d k_2 d m_1 d m_2 \, , \nonumber \\
R^k_{12}=\Delta^{-1}_{k 1 2} \,
   \delta(\omega_{p}-\omega_{p_1}-\omega_{p_2}) \,
f^k_{12} \, |V^k_{12}|^2 \, \delta_{m-m_1-m_2} k k_1 k_2
\, , \nonumber \\
 \Delta^{-1}_{k 1 2} = \left< \delta({\bf k}-{\bf
k_1}-{\bf k_2})\right>\equiv \int \delta({\bf
k}-{\bf k_1}-{\bf k_2}) \, d \theta_1 d \theta_2 \, , \nonumber\\
\Delta _{k 1 2} = \frac{1}{2}\sqrt{
2 \left( (k k_1)^2 +(k k_2)^2 +(k_1 k_2)^2
\right)-k^4-k_1^4 -k_2^4} \, .
\end{eqnarray}
\section{Kolmogorov Spectra in the High Frequency Limit} 
\label{sec:Kol}

Once the kinetic equation is derived, it is natural to search for its
stationary solutions. Typically, kinetic equations admit two classes
of exact stationary solutions: thermodynamic equilibrium and
Kolmogorov flux solutions, with the latter corresponding to a direct
cascade of energy --or other conserved quantities-- toward the higher
modes.  The fact that the thermodynamic equilibrium --or equipartition
of energy-- $n_{\bf p} = 1/\omega_{\bf p}$ is a stationary solution of
(\ref{KEinternalAveragedAngles}) can be seen by inspection.  On the
other hand, the solutions, corresponding to pure Kolmogorov spectra
are much more subtle and only emerge after one has exploited scaling
symmetries of the dispersion relation and the coupling coefficient via
what is now called the Zakharov transformation~\cite{ZLF}.

In this section, we carry out such procedure in the high--frequency
limit, where the effects of the Earth's rotation are negligible.  The
spectrum that we find, expanding on previous work described in
\cite{LT}, is not far from the classical empirical formula of Garrett
and Munk.

Since the system under consideration has cylindrical, rather then
spherical symmetry, stationary solutions of the kinetic equation
should also have cylindrical symmetry. Things are further complicated
by the fact that our dispersion law (\ref{InternalWavesDispersion}) is
{\it not} scale invariant, and neither is the interaction matrix
element (\ref{MatrixElement}). However, in the high frequency limit
$\omega\gg f $, (\ref{InternalWavesDispersion}) becomes
$$ \omega_{\bf p}\equiv \omega_{{\bf k},m} \simeq
   \frac{g}{N \rho_0} \frac{k}{|m|} \, ,$$
Furthermore, in this limit, the matrix element (\ref{MatrixElement})
retains only it first term, $I^1_{23}$.  This is due to the fact that
the second $J^1_{23}$ and third $K^1_{23}$ terms are proportional to
$f ^2$ and $ f $ respectively, and $f$ is negligible in the high
frequency limit.

Indeed if one changes variables in (\ref{MatrixElement}) so that
$$\omega_i = N \xi_i,$$ rescaling the frequencies in terms of the
buoyancy frequency $N$, and similarly one introduces
$$\vec k_i = \vec \kappa_i / L,$$ i.e. nondimensionalizing 
the horizontal wavevectors
in terms of some distance $L$ to be determined, then 
\begin{eqnarray}
V^1_{23}&=& I^1_{23}+J^1_{23}+K^1_{23}\nonumber\\
I^1_{23}&=&-\frac{N^{\frac{3}{2}}}{4\sqrt{2 g}L} \left(
\frac{\kap_2\cdot\kap_3}{\kappa_2 \kappa_3}\sqrt{\frac{\xi_2
\xi_3}{\xi_1}} \kappa_1 + \frac{\kap_1\cdot\kap_3}{\kappa_1
\kappa_3}\sqrt{\frac{\xi_1 \xi_3}{\xi_2}} \kappa_2 +
\frac{\kap_1\cdot\kap_2}{\kappa_1 \kappa_2}\sqrt{\frac{\xi_1
\xi_2}{\xi_3}} \kappa_3 \right)
\nonumber\\
 J^1_{23}&=&\frac{N^{3/2}}{4 L\sqrt{ 2 g}}
\frac{f^2}{N^2} \frac{1}{\sqrt{ 2 g \ \xi_1 \xi_2 \xi_3}}
\left(
\frac{\kap_2\cdot\kap_3}{\kappa_2 \kappa_3}{\kappa_1}
-\frac{\kap_1\cdot\kap_3}{\kappa_1 \kappa_3}{\kappa_2}
-\frac{\kap_1\cdot\kap_2}{\kappa_1 \kappa_2}{\kappa_3} \right)
\nonumber \\ 
K^1_{23} &=& i\frac{N^{3/2}}{L  \sqrt{2 g}}\frac{f}{N}
\frac{\kap_2\cdot\kap_3^{{\perp}}}{\kappa_1 \kappa_2 \kappa_3}
\left(\sqrt{\frac{\xi_2}{\xi_1\xi_3}}(\kappa_1^2-\kappa_3^2) +
\sqrt{\frac{\xi_1}{\xi_2\xi_3}}(\kappa_2^2-\kappa_3^2) +
\sqrt{\frac{\xi_3}{\xi_1\xi_2}}(\kappa_2^2-\kappa_1^2) \right)
\nonumber
\end{eqnarray}
Note that $K^1_{23}$ is proportional to $(f/N)$ and $J^1_{23}$ is
proportional to $(f/N)^2$. Taking into account that, in the real
ocean, $f/N\simeq 1/100$ we see that the $K^1_{23}$ and $J^1_{23}$
terms could safely be neglected from the matrix element in the high
frequency limit.

We reproduce here for completeness the derivation of the Kolmogorov
solution found in~\cite{LT}. Let us assume that $n_{\bf p}$ is given
by the power-law anisotropic distribution
\begin{equation}\label{n}
  n_{{\bf k},m}= k^x |m|^y \, .
\end{equation}
To find the values of the exponents $x$ and $y$ we will require that
(\ref{n}) is a stationary solution to (\ref{KEinternalAveragedAngles})
in the high frequency limit.  We shall use a version of Zakharov's
transformation \cite{Z68a,Z68b} introduced for cylindrically
symmetrical systems by Kuznetsov in~\cite{KuznetsovConformal}.
Namely, let us subject the integration variables in the second term
$R^1_{k2}$ in (\ref{KEinternalAveragedAngles}) to the following
transformation:
\begin{eqnarray}\nonumber
k_1 = {k^2}/{k_1'}, \
m_1={m^2}/{m_1'}, \
k_2 = {k k_2'}/{k_1'}, \  m_2 = {m m_2'}/ {m_1'}.
\nonumber\end{eqnarray} Then $R^1_{k2}$ becomes $R^k_{12}$ multiplied
by a factor
$${\left(\frac{k_1}{k}
\right)^{-6 - 2 x}
\left(\frac{m}{m_1}
\right)^{2 + 2y}}.$$ 
Furthermore, let us subject the third term $R^2_{1k}$ in
(\ref{KEinternalAveragedAngles}) to the similar conformal
transformation:
\begin{eqnarray} \nonumber
k_1 = {k k_1' }/{k_2'},
 \   m_1={m m_1'}/{m_2'},
\
k_2 = {k^2}/{k_2'},  \ m_2 = {m^2}/{m_2'}.
\end{eqnarray}
Then  $R^2_{1k}$ becomes $R^k_{12}$ multiplied by a factor
$$\left(\frac{k_2}{k}\right)^{-6 - 2 x}
\left(\frac{m}{m_2} \right)^{2 + 2y}.$$
Therefore   (\ref{KEinternalAveragedAngles}) can be written as
\begin{eqnarray} \label{KEtransformed}\frac{d n_{\bf p}}{d t}
= \frac{1}{k}\int R^k_{12} \,
\left(1-
\left(\frac{k_1}{k}\right)^{-6 - 2 x}
 \left(\frac{m}{m_1} \right)^{2 + 2y}-\right. \nonumber \\ \left.
\left(\frac{k_2}{k}\right)^{-6 - 2 x}
\left(\frac{m}{m_2} \right)^{2 + 2y}
\right) \,  d k_1 d k_2 d m_1 d m_2 \, . \label{KEZtransformed}
\end{eqnarray}
We see that the particular choice $ -6- 2 x = 2 + 2 y = 1 \, ,$
which gives
$x=-{7}/{2}, \ \ \ \ y=-{1}/{2} \, ,$
makes the right-hand side of
(\ref{KEinternalAveragedAngles}) vanish due to the delta function
in the frequencies, corresponding to energy conservation.
The resulting wave action and spectral energy
distributions are given by
\begin{eqnarray} n_{{\bf k},m}= |{\bf k}|^{-7/2}
 |m|^{-1/2}, \ \ \\
E_{{\bf k},m} = k \omega_{{\bf k},m} n_{{\bf k},m}= |{\bf
k}|^{-3/2} |m|^{-3/2},\nonumber\end{eqnarray}
This solution corresponds to the flux of energy from the large to the
small scales.  Direct calculations show that this solution is {\it
local}, i.e. collision integral converges on this solution.

We shall next compare the spectrum just derived with the classic
formula by Garrett and Munk \cite{GM72,GM75,GM79}, which synthesizes
in compact form many measurements of oceanic internal wave turbulence.

To this end, let us first write the total energy of the system as the
integral of a spectral energy density:
\begin{eqnarray}
E=\int E({\bf k},m) \, d { k} \, d m  \, .
\end{eqnarray}
Note that $E({\bf k},m) $ is integrated over {\it scalars} $k$ and
$m$.  Garrett and Munk proposed the following empirical expression for
$E(k,m)$:
\begin{eqnarray}\label{GM}
  E(k,m)= \frac{3\, f \, N \, E  \,
     m/m^*}{\pi \left(1+\frac{m}{m^*}\right)^{5/2} \left(N^2 k^2 + f^2
     m^2\right)} \, .
\end{eqnarray}
Here $E$ is a constant, quantifying the total energy content of the
internal wave spectrum, $k=|{\bf k}|$, and $m^*$ is a reference
vertical wavenumber determined from observations.  Using the
dispersion relation for internal waves, the GM spectrum can be
transformed from wave-number space $(k,m)$ into frequency-horizontal
wavenumber space $(k,\omega)$.  The integral of $E(k,\omega)$ over $k$
yields the moored spectrum
\begin{equation}
  E(\omega) =
   {2\, f\, E}\left(
   {\pi \left(1-({f}/{\omega})^2\right)^{1/2}\omega^2}
\right)^{-1}
 \, ,
   \label{moored}
\end{equation}
with an $1/\omega^2$ dependence away from the inertial frequency that
appears prominently in moored observations.

For $|m| >> m^*$ and $|\omega| >> f$, the Garrett-Munk spectrum
(\ref{GM}) becomes
\begin{equation}
E(k, m) \simeq \left(k^2 m^{3/2}\right)^{-1} \, . \label{GMLW}
\end{equation}
By contrast, the spectrum that we obtained above using the Wave
Turbulence formalism is
\begin{equation}
E(k, m) \simeq \left( k^{3/2} m^{3/2} \right)^{-1} \, . \label{GMWT}
\end{equation}

The small difference between the two spectra may be due to a variety
of reasons. Two possibilities that we find highly plausible are the
neglected effects of the Earth's rotation, and those of breaking
waves.  An exploration of these possibilities, however, lies outside
the scope of the present article.

\section{Conclusions}
\label{sec:concl}

We have developed a quite general, natural Hamiltonian formalism for
internal waves in a stratified, rotating environment. Our formulation
gains much in simplicity, by restricting consideration to flows in
hydrostatic balance, superimposed on a vertically arbitrary, but
horizontally uniform shear and vorticity fields.  The resulting
Hamiltonian inherits much of the structure of the shallow--water
equations, though with one extra vertical dimension.  The use of
isopycnal coordinates, whereby the depth $z$ is replaced by the
density $\rho$ as the independent vertical coordinate, allows for a
straightforward separation of the dynamics of waves and vorticity, by
assuming the latter to be uniform on surfaces of constant density.

This Hamiltonian formulation allows us to derive a kinetic equation
for the time evolution of the spectral energy density.  In the limit
of high frequencies, when the effects of the rotation of the Earth
loose significance, an exact steady solution to this kinetic equation
can be found, corresponding to the direct cascade of energy toward the
short scales. This Kolmogorov--like spectrum is surprisingly close to
the empirically based prediction of Garrett and Munk.

Further challenges suggested by the work reported here, include
extending the Kolmogorov solutions found in the high-frequency limit,
to cover the full range of frequencies of internal waves, including
those comparable to the inertial frequency $f$. Also, the effects of
breaking waves on the energy spectrum needs to be addressed; it could
potentially help explain the difference between the exponent of the
spectrum found here and that of Garrett and Munk.

\end{document}